   \documentstyle[proceedings,psfig]{crckapb} 

\begin{opening}

\title{THE ROLE OF NUCLEAR STARBURSTS IN LUMINOUS INFRARED SEYFERT 2 GALAXIES}

\author{ROSA M. GONZ\'ALEZ DELGADO}
\institute{Instituto de Astrof\'\i sica de Andaluc\'\i a\\
           Apdo. 3004, 18080 Granada, Spain}
\author{TIMOTHY HECKMAN}
\institute{Department of Physics and Astronomy, JHU\\
           Batimore, MD 21218, USA}

\end{opening}

\runningtitle{NUCLEAR STARBURST IN SEYFERT 2 GALAXIES}

\begin{document}


\begin{abstract}
We present HST (WFPC2 and FOC) images and UV GHRS spectra
plus ground-based optical spectra of four Luminous Infrared Galaxies (LIRGs) 
that have Seyfert 2
nuclei (Mrk 477, NGC 7130, NGC 5135 and IC 3639). 
The data provide direct evidence of the existence of a central nuclear
starburst that dominates the UV and optical light and are dusty and 
compact. The bolometric
luminosity ($\sim$ 10$^{10}$ L$\odot$) of these starbursts is similar to the estimated 
bolometric luminosities of their obscured  Seyfert 1 nuclei, and thus they contribute
in the same amount to the overall energetics of these galaxies.  An extended work based on ground-based 
optical spectra of the 20 brightest nuclei known indicate that at least 40$\%$ of the 
Seyfert 2 galaxies harbor a nuclear starburst. The eight Seyfert 2 nuclei that harbor a starburst are strong IR emitters. 
This suggests that nuclear starbursts can make a significant contribution or even dominate the UV 
and optical light of LIRGs.

\end{abstract}

\section{Introduction}

Ultraluminous Infrared Galaxies (ULIRGs), even being rare objects, are the 
dominant population in the Local Universe. They may have formed by the interactions 
and mergers of molecular gas-rich spirals, that emit the maximum of their 
infrared luminosity when the two nuclei merge. They may also represent the 
first stage in the formation of the cores of elliptical galaxies (Sanders \& 
Mirabel 1996). Theoretical models show that the merger process can accumulate 
a large concentration of gas in the centers of these galaxies (Barnes \& 
Hernquist 1992). This central reservoir of gas can fuel a powerful Starburst 
or an Active Galactic Nucleus (AGN), one of these two phenomena being the 
dominant energy source in ULIRGs. In fact, the nuclear spectrum of 2/3 of these 
galaxies can be 
classified as Seyfert or Liner and the remaining 1/3 as Starburst  galaxies 
(Kim 1995). However, there are also some galaxies where the Starburst and 
AGN characteristics co-exist in their nucleus. One of the famous ULIRGs is Arp 220 
(Londsdale et al 1994; Smith et al 1998). Most of these ULIRGs are very far 
away, and the observations do not have enough spatial resolution to measure independently
the contribution of the Starburst and AGN components to the total luminosity
of these galaxies. 

To shed some light on the question of the nature of the dominant energy source 
in infrared galaxies, we present here ultraviolet and optical data of 
four Seyfert 2 nuclei (Mrk 477, NGC 5135, NGC 5135 and IC 3639) that have IR 
luminosities $\sim$ 10$^{11}$ L$\odot$ and can be classified as Luminous 
Infrared Galaxies (LIRGs). They are located at a distance of $\sim$ 50 Mpc 
(except Mrk 477 which is at 150 Mpc). They are close enough that their nuclear 
zone can be 
resolved by the Hubble Space Telescope (HST) and to enable us to measure the 
Starburst and AGN contribution to the nuclear activity. This makes these Seyfert 2 galaxies benchmarks to study the
Starburst-AGN connection in more distant ULIRGs.
A more extended analysis of 
these data is presented in Heckman et al (1997) and Gonz\'alez Delgado et al 
(1998a). Here, this analysis is complemented with the results obtained from 
ground-based optical spectra of a sample of Seyfert 2 galaxies.

\section{The Sample}

The sample contains 20 of the brightest Seyfert 2 nuclei selected 
from the compilation of Whittle (1992). 
The brightness of the nucleus is defined as a function of the 
[OIII] $\lambda$5007+4959 emission line flux
and the monochromatic flux ($\nu \times F_{\nu}$)
of the nuclear non-thermal radio source at 1.4 GHz.
All of them satisfy at least one of the two following criteria: log
F$\rm_{[OIII]}\geq$ -12.0 (erg cm$^{-2}$ s$^{-1}$) and log F$_{1.4}\geq$
-15.0 (erg cm$^{-2}$ s$^{-1}$). The criteria used guarantees that the sample 
is unbiased with respect to the presence or absence of a nuclear starburst. 

We have taken ground-based optical and near infrared spectroscopy of all
the targets with the 4m telescope at Kitt Peak National Observatory.
HST+WFPC2 archive images at optical wavelengths are also available for most of the objects. 
HST UV images have been obtained 
for 12 of the 20 Seyfert 2
nuclei using the Faint Object Camera (FOC). We have also obtained
UV spectra of 4 of these 12 Seyfert 2 nuclei with the HST Goddard
High Resolution Spectrometer (GHRS). These are: Mrk 477, NGC 7130, 
NGC 5135 and IC 3639. They were chosen from the subsample of 
12 Seyfert 2 nuclei for having the highest UV flux on arcsec scales. 

\section{UV HST imaging results}

UV images were obtained with the FOC using the F/96 relay through the filter
F210M. This filter does not 
include any strong emission lines; therefore, the emission is dominated by the
UV continuum at 2150 \AA. The field of view is about 14$\times$14
arcsec and the pixel size 0.014 arcsec. 

In the four galaxies, the UV images show that the UV continuum source is spatially resolved. 
The central 1 arcsec (730 to 210 pc) shows a morphology very similar to starburst galaxies. 
The effective radius of the starburst that ranges from 55 pc in IC 3639 to 200 pc in NGC 5135 
indicates that they are very compact. They are formed by several knots that have sizes and 
luminosities very similar to young clusters (Meurer et al 1995). The optically
bright nucleus is barely detected at the UV (see other examples in Colina et al 1997), and their UV 
emission is only a few percent ($\leq$ 15$\%$) of the nuclear starburst emission (Figure 1).

\begin{figure}


\psfig{file=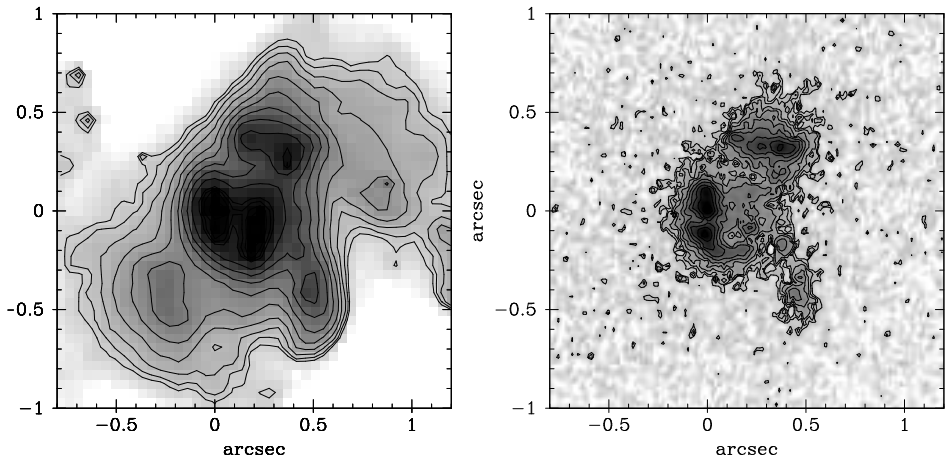,height=6cm}

\caption[fig]{ HST/WFPC2 (F606W) (left) and HST/FOC (F210M) (right) image of the
nuclear region (2$\times$2 arcsec equivalent to 620 by 620 pc) of the
Seyfert 2 LIRG NGC 7130. The UV image clearly shows a very compact
starburst resolved in several knots. The presumed nucleus is placed at 0.2
arcsec West and 0.1 arcsec South of the origin. This position in the
optical image shows a strong emission  probably due to emission line gas
and continuum from older stars.}

\end{figure}

\section{UV HST spectroscopy results}

The four galaxies were observed with the GHRS and the G140L grating,
which has a nominal dispersion of 0.57 \AA/diode, using the Large Science
Aperture (LSA, 1.74$\times$1.74 arcsec) and covering 1150-1600 \AA. The spectra show
 the typical absorption features of starburst galaxies, indicating 
that the UV light
is dominated by the starburst component. However, they also show narrow
emission lines typical of Seyfert 2 galaxies (L$\alpha$ and CIV $\lambda$1549, NIV] $\lambda$1486, SiIV+OIV] $\lambda$1397+1403, NV $\lambda$1240, and CII$\lambda$1335). 

The spectra are very rich in absorption
features formed in the photospheres of O and B massive stars and in the interstellar medium of the galaxies. The interstellar lines are
blueshifted by several hundred km s$^{-1}$ with respect to the systemic velocity of the galaxies, indicating that the warm interstellar gas in these
galaxies is outflowing. Outflows of similar velocity have been
detected in starburst galaxies (Gonz\'alez Delgado et al 1998b).  However, the strongest absorption features are the resonance lines CIV $\lambda$1550, 
SiIV $\lambda$1400 and NV $\lambda$ 1240 formed in the stellar wind of O stars. These lines are very prominent if the spectrum 
is dominated by a starburst with age between 1 to 10 Myr.  Evolutionary synthesis models 
based on the profile of the stellar wind and UV continuum luminosity indicate that these starbursts are heavily
reddened (2 or 3 mag in the UV), very powerful (bolometric luminosities of $\sim$ 10$^{10}$ L$\odot$),
were of short duration and have ages between 4 and 6 Myr (Figure 2a).

\begin{figure}


\psfig{file=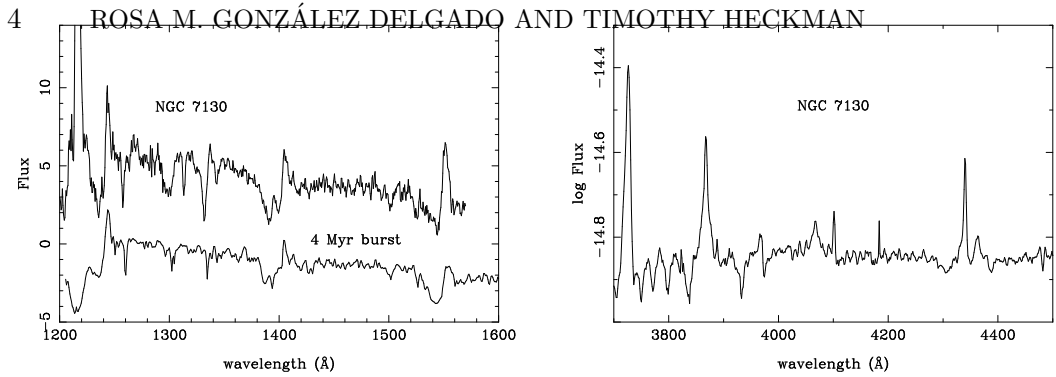,height=4.5cm}

\caption[fig]{(a) GHRS spectrum of NGC 7130 dereddened  and the synthetic 4 Myr burst model (in relative units). The IMF slope is Salpeter and M$_{upp}
$=80 M$\odot$. (b) Ground-based optical spectrum of NGC 7130 obtained through a 1.5$\times$3.5 arcsec$^2$ aperture.}

\end{figure}

\section{Ground-based optical spectroscopy}

Signatures of massive stars are also clearly detected in their
optical and near-UV spectra where the high order Balmer series (H9, H10, and H11) and 
HeI lines ($\lambda$4921, $\lambda$4387, $\lambda$4026 and $\lambda$3819)
are observed in absorption (Figure 2b). These lines are formed in the photospheres of
O and B stars, and thus they also provide strong independent
evidence of the presence of massive stars in these Seyfert 2 nuclei.  They
have not been detected in Mrk 477 because the Balmer-emission lines formed in the Narrow Line 
Region are so strong that they overwhelm the stellar absorption lines. However, its spectrum shows 
a broad emission line around HeII $\lambda$4686 that may be produced by Wolf-Rayet stars. 
In the near-infrared we detect a strong CaII triplet  feature produced by red supergiant stars.

This detailed study shows that when a conspicuous starburst is present in the nucleus of Seyferts, the UV image shows a morphology similar to starburst galaxies, the UV spectrum shows 
strong wind P-Cygni profiles, and the ground-based optical spectrum shows the high-order 
Balmer series and HeI in absorption. These results thus prove that the high-order Balmer series and HeI lines constitute a good indicator of the presence of massive stars in Seyfert 
nuclei. The analysis of the whole sample of Seyfert 2 galaxies suggests that a starburst 
is present in 40$\%$ of the nuclei (one example is in Figure 3b) because these lines are detected in the nuclear spectrum of eight of the twenty galaxies of the sample. On the other hand, the HST optical archive images of these galaxies show a nuclear morphology very similar to starburst galaxies, like 
the four Seyfert 2 galaxies studied by us in detail (Figure 3a).

\begin{figure}


\psfig{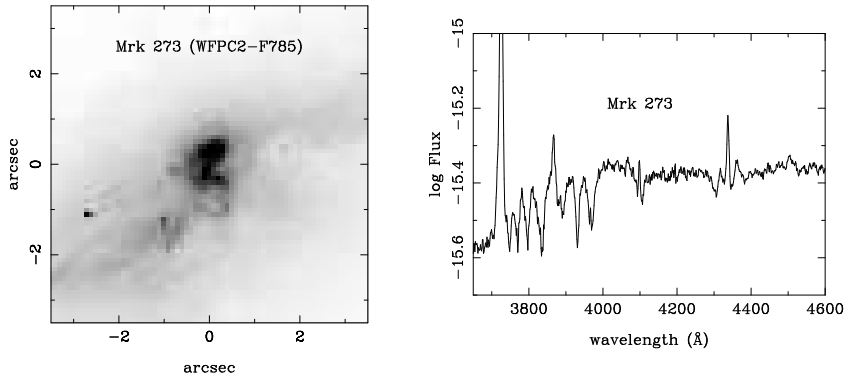}

\caption[fig]{(a)HST/WFPC2 (F785W) image of the
central region (1 arcsec equivalent to 730 pc) of the
ULIRG Seyfert 2  Mrk 273. (b) Ground-optical spectrum of Mrk 273 obtained through a 1.5$\times$3.5 arcsec$^2$ aperture.}

\end{figure}

\section{Conclusions and Implications for LIRGs}

The data provide direct evidence of the existence of nuclear starbursts that dominate the 
ultraviolet and optical light in at least 40$\%$ of the Seyfert 2 galaxies. The starburst is heavily 
reddened, compact and powerful. The bolometric luminosities of these starbursts are very similar 
to the estimated bolometric luminosities of their obscured Seyfert 1 nuclei. These data suggest 
that more powerful AGNs may be related to more powerful central starbursts. These starbursts also make  a significant contribution to the overall energetic of these galaxies. 

The 40$\%$ of the  Seyfert 2 galaxies of our sample that show in their UV and/or optical nuclear spectra signatures of 
massive stars have FIR luminosities over 10$^{44}$ erg s$^{-1}$, even one of them, Mrk 273, has IR luminosity up to 10$^{12}$ L$\odot$; thus, they all can be classified as LIRGs. There are other 4 galaxies in our sample that are very strong IR emitters but they do not show massive stellar features in their ground-based optical spectra. There are WFPC2 or FOC images for three of these, and two of them 
show Starburst morphology. It may be that in these two cases, the starburst is less conspiscuous than in the other Seyfert 2 nuclei and they are not detected from the ground. Therefore, we can conclude that 
most of the LIRGs in our sample harbor a nuclear starburst that makes a significant contribution to or even dominates the energetics  of a significant fraction of the LIRGs. However, the exception in our sample is Mrk 463E that does not show a starburst morphology at UV wavelengths, and the optical spectrum does not show massive stellar features. On the contrary, its UV FOC image shows a conical diffuse structure that could be associated with the ionizing cone of the obscured Seyfert 1 nucleus. Broad emission Balmer components have been detected in polarized light (Tran 1995). It may that in this LIRGs, the main energy source is a AGN.

\end{document}